\documentclass[prl,showpacs,floatfix,twocolumn]{revtex4}
\usepackage{epsfig}

\begin{document} 
\title{Smearing of charge fluctuations in a grain by spin-flip 
assisted tunneling}

\author{Karyn Le Hur$^1$ and Pascal Simon$^{2,3}$}
\affiliation{$^1$ D\'epartement de Physique and CERPEMA,
 Universit\'e de Sherbrooke, Sherbrooke, Qu\'ebec, Canada, J1K 2R1}
\affiliation{$^2$ Department of Physics and Astronomy, University of
Basel, Klingelbergstrasse 82, 4056 Basel, Switzerland\\
$^3$ Department of Physics, Boston University, 590 Commonwealth av., 
Boston, MA02215}

\newcommand{\br}{{\bf r}}
\newcommand{\ovl}{\overline}
\newcommand{\hw}{\hbar\omega}
\newcommand{\mybeginwide}{
    \end{multicols}\widetext
    \vspace*{-0.2truein}\noindent
    \hrulefill\hspace*{3.6truein}
}
\newcommand{\myendwide}{
    \hspace*{3.6truein}\noindent\hrulefill 
    \begin{multicols}{2}\narrowtext\noindent
}
 
\date{\today} 
\begin{abstract}
We investigate the charge fluctuations of a grain coupled to a lead
via a small quantum dot in the Kondo regime. We show that the 
strong entanglement
of charge and spin flips in this setup can result in a 
{\it stable} 
SU(4) Kondo fixed point, which considerably smears out 
the Coulomb staircase behavior 
already in the weak tunneling limit. This behavior is robust enough
to be  experimentally observable. 
\end{abstract}

\maketitle

Recently, quantum dots have attracted a considerable interest due to their potential
applicability as single electron transistors or as basic building blocks 
(qubits) in the fabrication of quantum computers \cite{Loss}. One of 
their most important features 
is the Coulomb blockade phenomenon, i.e., as a result of the
strong repulsion between electrons, the charge of a quantum dot
is quantized in units of the elementary charge $e$. Even when
the quantum dot is weakly-coupled to a bulk lead, so that electrons
can hop from the lead to the dot and back, the dot charge remains to a large 
extent
quantized. This quantization has been thoroughly investigated 
both theoretically \cite{Matv1} and experimentally \cite{Grabert}. The 
quantity of interest here is
the average dot charge as a function of the voltage applied to a back-gate.
For a weakly-coupled dot, at low temperatures, 
this function generally exhibits sharp steps, 
resulting in the ``Coulomb staircase''. We emphasize the fact that
a direct measurement of the average dot charge can be performed
with  sensitivity well below a single charge \cite{Berman}.

In this Letter, we investigate the shape of the steps of the
Coulomb staircase in the presence of spin-flip assisted
tunneling.
The setup we examine consists of a (large) dot or grain coupled to a
reservoir through a smaller dot (Fig.1). We assume that the smaller dot 
contains
an odd number of electrons and acts 
as an  S=1/2 Kondo impurity \cite{Glazman}.  
A different limit where
the small dot rather acts as a resonant level has been partially studied in 
Ref. \onlinecite{Gramespacher} where it was shown that the resonant level
can smear the Coulomb blockade (nevertheless, for a narrow resonance at the
Fermi energy, the effect is weak). 
Furthermore, the charge of the grain in such a device can be used to measure 
the occupation of the dot \cite{Gramespacher}.

\begin{figure}[ht]
\centerline{\epsfig{file=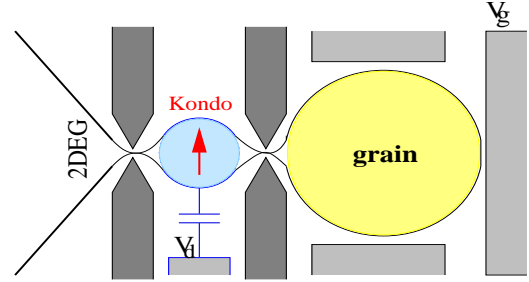,angle=0.0,height=6.cm,width=
6.85cm}}
\vskip -2.5cm
\caption{Schematic view of the setup. A grain (large dot) with a micronscale 
is weakly-coupled to a bulk
lead via a small dot in the Kondo regime which acts as an S=1/2 spin impurity.
The charges of the grain and the small dot are  controlled by the gate voltage 
$V_g$ and $V_d$ respectively.
The auxiliary voltages can be used to adjust the (symmetric) tunnel junctions.}
\end{figure}

In the following analysis, we focus on the  regime where the
back-gate voltage is such that the charging states of the grain with
0 or 1 electron are {\it degenerate}. 
We show below that in this case, the charge
degrees of freedom of the grain become strongly entangled with the spin degrees
of freedom of the small dot resulting in a  stable fixed point with an SU(4)
symmetry. The major consequence of this enlarged symmetry in our
setup is that the dot's capacitance exhibits instead of a logarithmic 
singularity \cite{Matv1}-
a strongly reduced peak as a function of the back-gate voltage, 
smearing  charging effects in the grain considerably. The
Coulomb staircase behavior 
becomes  smeared out already in the weak tunneling limit due to the prominence
of spin-flip assisted tunneling. The possibility of a strongly correlated 
ground state possessing an SU(4) symmetry has also been  discussed very
recently 
in the (very) 
different context of two small dots coupled with a strong capacitive
inter-dot coupling \cite{Zarand}.

The model we consider is described by the Anderson-like Hamiltonian:
\begin{eqnarray}
\label{anderson}
H&=&\sum_k \epsilon_k a^\dag_{k\sigma} a_{k\sigma}+
\sum_p \epsilon_p a^\dag_{p\sigma} a_{p\sigma} + 
{\hat Q^2\over 2C}+\varphi \hat Q \nonumber
\\ 
&+& \sum_{\sigma} \epsilon a_{\sigma}^{\dag}a_{\sigma}+U 
n_{\uparrow}n_{\downarrow}
\\ \nonumber
&+& t\sum_{k\sigma}(a^\dag_{k\sigma}a_{\sigma}+h.c.)+
t\sum_{p\sigma}(a^\dag_{p\sigma}a_{\sigma}+h.c.),
\end{eqnarray}
where $a_{k\sigma}$, $a_{\sigma}$, $a_{p\sigma}$ are the annihilation
operators for electrons of spin $\sigma$ in the lead, the small dot, and
the grain, respectively, and $t$ is the tunneling matrix element 
which is assumed not to depend on $k$.
We assume in the sequel that the junctions are {\bf symmetric}  
and narrow enough, i.e., contain {\it one} channel only.
The energy spectrum in
the grain is assumed to be {\it continuous}, which implies that the grain is
sufficiently large (at the micron scale\cite{Berman}). 
$\hat Q$ denotes the charge operator of the grain, 
$C$ is the capacitance between the grain and the gate electrode, and $\varphi$
is related to the back-gate voltage $V_g$ through $\varphi=-V_g$. 
$\epsilon<0$ and $U$ 
stand for 
the energy level and charging energy of the small dot, and $n_{\sigma}=
a_{\sigma}^{\dag}a_{\sigma}$. The inter-dot capacitive coupling is weak and 
neglected here.
We consider the situation where the small dot is in the Kondo regime
(which requires  the last level to be singly occupied and 
the condition
$t\ll (-\epsilon,U+\epsilon)$ to be satisfied). 

After a standard Schrieffer-Wolff transformation\cite{Hewson}, 
the system is described by the Hamiltonian:
\begin{eqnarray}
\label{ham}
H&=&\sum_{k} \epsilon_k a^\dag_{k} a_{k}+
\sum_p \epsilon_{p} a^\dag_{p}a_{p} + 
{\hat Q^2\over 2C}+\varphi \hat Q\\ \nonumber
&+&\sum\limits_{m,n}({J\over 2}\vec S\cdot{\vec{\sigma}}+V)a_{m}^\dag 
a_{n}.
\end{eqnarray}
To simplify notation, the spin indices have been omitted and
hereafter.
$m,n$ take values in the two
sets ``lead'' (k) or ``grain'' (p), the spin $\vec S$ is the spin 
of the small dot, $\vec{\sigma}$ are Pauli matrices acting on the spin space
of the electrons and 
$J=2t^2[1/(-\epsilon)+1/(U+\epsilon)]$ is the related 
Kondo coupling. A direct hopping term $V
=t^2/2[1/(-\epsilon)-1/(U+\epsilon)]$ is also present. We have neglected the
charging energy of the grain $E_c=e^2/(2C)\ll (|\epsilon|,U)$. 

We would like to compute the corrections to the
average charge on the grain due to the $J$ and $V$ couplings 
at {\it zero} temperature bearing in mind that when the tunneling amplitude 
$t\rightarrow 0$, i.e., 
$J=0$ and $V=0$, the average grain
charge exhibits  perfect Coulomb staircase behavior as a function of $V_g$.
We confine ourselves to values of $\varphi$ in the range 
$-e/(2C)\leq\varphi<e/(2C)$,
which corresponds to the unperturbed (charge) value $Q=0$. 
A first intuitive approach is to assume that $(J,V)$ 
are very small compared to the charging energy $E_c=e^2/(2C)$ of the grain
and to calculate the corrections to $Q=0$ in perturbation theory. At second
order in perturbation theory, we find
\begin{equation}
\langle\hat{Q}\rangle\ =e({3\over 8} J^2+2V^2)\ln\hbox{\huge{(}}{e/2C-
\varphi\over
e/2C+\varphi}\hbox{\huge{)}}.
\end{equation}
The density of states in the lead and in the grain have been assumed to be 
equal\cite{Matv1} and taken to be
1 for simplicity. 
This result  generalizes 
that of a grain directly coupled to a lead \cite{Matv1}.

There are two reasons that may suggest this perturbative approach is divergent.
Higher-order terms (already at cubic orders) involve logarithmic divergences
associated with the Kondo coupling, but also other logarithms
indicating the vicinity of a degeneracy point in the charge sector. Note that 
the (degeneracy) point where the grain charging states with $Q=0$ and $Q=e$
are degenerate corresponds explicitly to $\varphi=- e/2C$.
Below, we will be primarily interested in the situation close to this 
degeneracy point, i.e., {\it in the height of the capacitance peaks} at low
temperature, where none of the perturbative arguments above can be applied.
The Hamiltonian (\ref{ham}) can be mapped onto some 
Kondo Hamiltonian following  Ref. \cite{Matv1}. 
We introduce the
projectors $\hat P_0$ and $\hat P_1$ which project on the  states with $Q=0$ 
and
$Q=e$ in the grain. The truncated Hamiltonian (\ref{ham}) then reads:
\begin{eqnarray}
H&=&\sum\limits_{k,\tau=0,1} \epsilon_k 
a_{k\tau}^\dag a_{k\tau}(\hat P_0+\hat P_1)+eh\hat P_1\\ 
\nonumber
&+& \sum\limits_{k,k'}
\hbox{\Large{[}}
\hbox{\large{(}} {J\over 2}{\vec \sigma}\cdot\vec S+V
\hbox{\large{)}} (a_{k1}^\dag
a_{k'0}\hat P_0+
a_{k'0}^\dag a_{k1}\hat P_1)
\\ \nonumber
&+&\sum_{\tau=0,1} \hbox{\large{(}} {J\over 2}{\vec\sigma}\cdot\vec
S+V\hbox{\large{)}} a_{k\tau}^\dag a_{k'\tau}\hbox{\Large{]}},
\end{eqnarray}
where now the index 
$\tau=0$ indicates the reservoir and $\tau=1$ indicates the grain.
We have also introduced the small parameter $h=e/(2C)+
\varphi\ll e/C$ which measures deviations from the degeneracy point.
Considering $\tau$ as an abstract {\it orbital} index, the Hamiltonian can be
rewritten in a more convenient way by introducing another set of Pauli
matrices for the orbital sector \cite{Matv1,Georg}:
\begin{eqnarray}
H&=&\sum\limits_{k,\tau} \epsilon_k a_{k\tau}^\dag a_{k\tau}-ehT^z 
\\ \nonumber
&+& \sum\limits_{k,k'}\hbox{\Large{[}}
\sum\limits_{\tau,\tau'}
\hbox{\large{(}}
{J\over 2}{\vec\sigma}\cdot\vec S+V\hbox{\large{)}}
(\tau^x T^x+\tau^yT^y)_{\tau,\tau'}a_{k\tau}^\dag a_{k'\tau'}
\\ \nonumber
&+ &\hskip 0.7cm 
\sum_\tau\hbox{\large{(}} 
{J\over 2}{\vec\sigma}\cdot\vec S+V\hbox{\large{)}} 
a_{k\tau}^\dag a_{k'\tau}
\hbox{\Large{]}}.
\end{eqnarray}
In this equation, the operators $(S,\sigma)$ act on 
spin and the $(T,\tau)$ act on the (charge) orbital
degrees of freedom. Here, $h$ mimics a magnetic field 
acting in orbital space. An important
consequence of this mapping is that the average grain charge 
$\langle\hat{Q}\rangle$ can be
identified as
$
%\begin{equation}\label{charge}
\langle\hat{Q}\rangle = 
e ({1\over 2} - \langle T^z \rangle ).
%\end{equation}
$
The grain capacitance 
$C=-\partial\langle\hat{Q}\rangle/\partial h$ is thus equivalent to the
local susceptibility $\chi_T=\partial\langle T^z \rangle/\partial h$.
% up to a factor of $e$.
Naturally, to compute the latter, we have to determine 
the nature of the Kondo ground state exactly.

Typically, when only ``charge flips'' are involved through the $V$ term, 
the model can be mapped onto a two-channel Kondo model, and 
the capacitance always exhibits
a {\it logarithmic divergence} \cite{Matv1}. Here, we have a 
combination of spin and charge flips. Can we then expect two distinct energy
scales for the spin and orbital sectors? To answer this 
question, we perform a perturbative scaling analysis following that of a
related model in Ref. \cite{Zarand2}. We first 
rewrite the interacting part of the Hamiltonian in  real space as: 
\begin{eqnarray}\label{heff}
H_K &=& {J\over 2}\vec{S}\cdot(\psi^{\dag}{\vec{\sigma}}\psi) \nonumber
\\ 
&+& {V_z\over 2}T^z(\psi^{\dag}
{\tau^z}\psi)+{V_{\perp}\over 2}[T^+(\psi^{\dag}{\tau^-}\psi)+h.c.]
\\ \nonumber
&+& Q_z T^z\vec{S}\cdot(\psi^{\dag}{\tau^z}{\vec{\sigma}}\psi)
+ Q_{\perp}\vec{S}\cdot[T^+(\psi^{\dag}{\tau^-}{\vec{\sigma}}\psi)+h.c.],
\end{eqnarray}
where $\psi_{\sigma\tau}=\sum_k a_{k\sigma\tau}$, $J$ refers to pure
spin-flip processes involving the S=1/2 spin of the small dot, 
$V_{\perp}$ to pure charge flips which modify the grain charge, and 
$Q_{\perp}$ describes exotic 
spin-flip assisted tunneling. 
This Hamiltonian exhibits a structure which is very similar
to the one introduced by Borda {\it et al.} in order to study
a symmetrical double (small) quantum dot structure 
with strong capacitive coupling \cite{Zarand}. However, since the physical situation
that led us to this Hamiltonian is very different from Ref. \cite{Zarand}, 
our bare values for the coupling parameters are also very 
different (for $J\ll 1$ and $V/J$ not too small, i.e., $2\epsilon+U$
not too close to 0): 
\begin{equation}
\label{couplings}
V_{\perp}=V,\ V_z=0\ ,\ Q_z=0\ ,\ Q_{\perp}=J/4.
\end{equation}
We have ignored the  potential scattering $V\psi^{\dag}\psi$ which
does not renormalize at low energy.

By integrating out conduction electrons with energy larger than an energy scale
$E\ll \Delta_d$ ($\Delta_d$ being the level spacing of the 
small dot, i.e., the ultraviolet cutoff), we 
obtain at second order 
the following renormalization group (RG) equations for the five dimensionless 
coupling 
constants:
\begin{eqnarray} \label{RG}
\frac{d J}{dl} &=& J^2 + {Q_z}^2 +2{Q_{\perp}}^2 \nonumber \\
\frac{d V_z}{dl} &=& {V_{\perp}}^2 +3 {Q_{\perp}}^2  \nonumber \\
\frac{d V_{\perp}}{dl} &=& V_{\perp}V_z+3 Q_{\perp}Q_z \\ 
\frac{d Q_z}{dl} &=& 2JQ_z +2V_{\perp}Q_{\perp}  \nonumber\\
\frac{d Q_{\perp}}{dl} &=& 2JQ_{\perp} + V_zQ_{\perp}+ V_{\perp} Q_z,\nonumber
\end{eqnarray}
with $l=\ln[\Delta_d/E]$ being the scaling variable.
This RG analysis  
is  applicable only very close to the degeneracy point $\varphi=-e/(2C)$
where the effective Coulomb energy in the grain or $h$ vanishes and obviously
only when all coupling constants stay $\leq 1$.
Higher orders in the RG have been 
neglected. We have integrated  the RG
equations (\ref{RG}) numerically.
It is noteworthy that even though we started with completely {\it asymmetric}
bare values of the coupling constants, due to the presence
of the spin-flip assisted tunneling terms $Q_{\perp}$ and $Q_{z}$, all
couplings diverge at the {\it same} Kondo temperature 
scale (again, assuming the bare ratio $V/J$ not too small) 
$T_K\sim \Delta_d\ e^{-1/4J}$. For example, we have checked numerically that 
all coupling ratios  converge to 
{\bf one} in the low energy limit provided the RG equations can be 
extrapolated 
in this regime. 
Therefore, this RG analysis suggests that, as in  
Ref. \cite{Zarand},
our model becomes equivalent at low energy to an SU(4) {\it symmetrical} 
exchange model $(J\gg 1)$:
\begin{eqnarray} \label{irrrep}
H_K &=& J \sum\limits_{A} \psi^{\dag} t^A
\hbox{\Large{[}}\sum\limits_{\alpha\beta}
(S^{\alpha}+{1\over 2})(T^{\beta} +{1\over 2})\hbox{\Large{]}}^A\psi \\ \nonumber
&=&  {J\over 4}
\sum\limits_{A} M^A\sum\limits_{\mu,\nu} \psi^{\dag}_{\mu}
t^A_{\mu\nu}\psi_{\nu},
\end{eqnarray}
where we have introduced the ``hyper-spin'' $M^A\in \{2S^{\alpha},2T^{\alpha},
4S^{\alpha}T^{\beta}\}$ for $\alpha,\beta=x,y,z$. The operators $M^A$ can be regarded
as the 15 generators of the SU(4) group. 
Moreover, this conclusion is reinforced by the numerical renormalization group analysis 
(whose range of validity is broader than Eqs. (\ref{RG})) of a model analogous 
to Eq. (\ref{heff}) developed in Ref. \cite{Zarand}.
Notice that  the irreducible 
representation of SU(4) written in Eq. (\ref{irrrep}) has been used previously 
 for spin systems with orbital 
degeneracy \cite{Li}.
The electron operator $\psi$ now transforms under the fundamental 
representation of the SU(4) group, with generators $t^A_{\mu\nu}$  
$(A=1,...,15)$, and the index $\mu$ labels the four combinations of
possible spin and orbital indices. The emergence of such a
strongly-correlated SU(4) ground 
state here clearly reflects the 
strong entanglement between the {\it charge} degrees of freedom of
the {\it grain} and the {\it spin} degrees of freedom of the {\it small dot} 
at low energy
due to the prominence of spin-flip assisted tunneling.

The (one-channel) SU(N) Kondo model has been extensively studied in the 
literature (see, e.g., Ref.\cite{Bickers}). Mostly, the strong 
coupling regime 
corresponds to a {\it dominant Fermi liquid} fixed point induced by the
complete screening of the hyper-spin $M^a$, implying that 
all the generators of SU(4) yield a local susceptibility with a behavior in 
$\sim 1/T_K$\cite{Parcollet}. 
$T^z$ being one of these generators, we deduce
 that 
$\chi_T=\partial\langle T^z \rangle/\partial h$ and then the capacitance of the
grain $C=-\partial\langle\hat{Q}\rangle/\partial h$ evolve as 
$1/T_K$\cite{Parcollet}. Consequently, for
$h\ll e/C$,
we obtain a {\bf linear} 
dependence of the average grain charge as a function
of $V_g=-\varphi$:
\begin{equation}\label{chargegrain}
\langle\hat{Q}\rangle -e/2 =-eh/T_K=-e[{e\over 2C}+\varphi]/T_K.
\end{equation} 
{\it This is the main result of this paper}.
The hallmark of
the formation of the SU(4) Fermi liquid 
in our setup is clear. The (grain) capacitance
peaks are completely smeared out by the mixing of spin and charge flips and
 Matveev's logarithmic singularity \cite{Matv1} has been completely destroyed. 

We now discuss the robustness of our SU(4) liquid. First, $S^z$ and 
$T^z$ are marginal operators which guarantees the stability of the SU(4)
fixed point at 
finite magnetic field $B\ll T_K$ and not too close to 
the degeneracy
point $h\ll T_K$\cite{Zarand}.  
The SU(4) symmetry should 
be still robust for wider junctions characterized by $n>1$ 
transverse channels\cite{Parcollet,longv};  even when the 
transmission amplitudes
for the $n$ modes are equal (extending results of Ref. \onlinecite{Zarand3}, 
an SU(4) Kondo singlet would occur at the renormalized Kondo 
temperature scale $T_K[n]\approx \Delta_d\ e^{-n}e^{-1/4J}$). 

Applying a strong magnetic field $B\gg T_K$ inevitably
destroys the SU(4) symmetry. However, we expect 
the behavior of charge fluctuations close to the degeneracy points
to remain  qualitatively similar. Indeed, in a large magnetic field spin flips 
are suppressed, i.e., $Q_{\perp}=Q_z=J=0$, and the orbital degrees of freedom,
through $V_{\perp}$ and $V_z$, develop a standard one-channel Kondo model
(the electrons have only spin-up or spin down), which also results in 
a Fermi-liquid ground state with a linear dependence of the average grain 
charge as in Eq. (\ref{chargegrain}). Yet, the emerging Kondo temperature
will be much smaller, $T_K[B=\infty]\approx \Delta_d \ e^{-1/2V}$, which 
affects the slope in (\ref{chargegrain}). Finally, for $h\gg T_K$, the orbital 
Kondo effect is destroyed and the results of ref.~\onlinecite{Gramespacher} 
should apply\cite{longv}.

In the following paragraph, we briefly analyze the conductance of the 
Kondo dot-grain (KDG) system
by connecting a  lead to the right of the grain in Fig 1.
We therefore add the following term
to the Hamiltonian (\ref{anderson}):
\begin{equation}
 \hat{H} = \sum_{k'} \epsilon_{k'} a^\dag_{k'\sigma} a_{k'\sigma}+  
t'\sum_{k'\sigma}(a^\dag_{k'\sigma}a_{p\sigma}+h.c.),
\end{equation}    
where  
$a_{k'\sigma}$ are the annihilation operators for electrons in the right
lead. For a realistic geometry  the charge flips 
between the right lead and the grain are inevitably the most prominent ones
(since $J\propto t^2\ll t'$
and $\Delta_d\leq \langle \hat{Q}^2\rangle/2C$).
This leads to the complete
screening of the orbital spin $T^z$ already at a temperature
scale $T_K^o\sim E_c\ e^{-1/t'}\gg T_K$\cite{Matv1}, and to an underlying 
two-channel Kondo model:
\begin{equation}
 \hat{H}_K^c ={1\over 2}t'\vec{T}\cdot\psi^{\dag}\vec{\tau}\psi.
\end{equation}   
The two orbital states here correspond to
$\tau=1$ for the grain and $\tau=2$ for the right lead. 
The transmission between the grain and the right
lead becomes perfect and the associated conductance is $2e^2/h$. 
As a result, the renormalization of the 
charge flip term $V_{\perp}$ between the
grain and the left lead becomes completely {\it cut off} 
by the screening of the
orbital impurity. Therefore, the direct hopping term between the left lead and the grain 
remains small implying that even for $\varphi=-e/2C$, 
the conductance through the KDG 
structure should be (very small) proportional to
$2V^2e^2/h$ \cite{note2}. Nevertheless, the spin flip term $J$ will 
continue to flow to strong coupling, making the ratio $Q_{\perp}(l)/J(l)\ll 1$
at low energy. As a consequence, 
another {\it non trivial} two-channel Kondo model appears in the spin sector
and the SU(4) symmetry is explicitly  broken.
This emergence of {\it two} distinct
two-channel Kondo models, 
both in the orbital and in the spin sectors 
deserves further investigation \cite{longv}.

To summarize, we have determined exactly the shape of the steps of the Coulomb
staircase for a grain coupled to a bulk lead through a small quantum dot
in the Kondo regime. We have shed light on the 
possibility of a stable SU(4) Fermi liquid fixed point, where
a Kondo effect appears simultaneously both in the spin and the 
orbital sectors. This requires the condition $2\epsilon+U\neq 0$ on
the small dot so that the bare $V_{\perp}/J$ isnot too small.
% It is worth noting that 
%the SU(4) Kondo fixed point is not really sensitive to the strong
%{\it asymmetry} between the bare values of coupling constants (for example for 
%$V\ll J$). 
In fact, due to the importance of spin-flip assisted tunneling, 
{\it charge} degrees of freedom of the {\it grain} become entangled 
with the {\it spin} degrees of freedom of the {\it small dot}, which 
explains
this enlarged symmetry (see Eq.(\ref{irrrep})). This implies the 
destruction of the 
logarithmic peak present in the Coulomb blockade 
already in the 
weak-tunneling limit. Such unusual behavior of charge
fluctuations in the grain close to the degeneracy points 
should be  
observable via capacitance measurements assuming the level spacing of the grain
$\Delta_g$
is small enough, i.e., $\Delta_g/E_c\rightarrow 0$. $T_K$ should not be 
too far from the Kondo scale in the 
conductance experiments through a single small quantum dot ($\sim 1K$)
\cite{wdv}), 
and capacitance measurements can be performed much below $100mK$\cite{Berman}.
Clearly, this effect is far more pronounced than the smearing of the Coulomb
blockade by a resonant impurity level (e.g., when $\epsilon\rightarrow 0$,
which can be tuned via the gate voltage $V_d$ of the small dot) 
which remains weak even if the transmission
through the impurity at the Fermi energy is perfect \cite{Gramespacher}.
This should be even more striking by measuring the charge of the grain as a 
function of the two gate voltages $V_g$ and $V_d$ \cite{note3}.
More generally, as in Ref. \cite{Zarand}, these results 
bring novel insight on the realization of Kondo ground states
with SU(N) $(N\gg 2)$ symmetry at the mesoscopic scale.

The authors acknowledge D. S\'en\'echal, A.-M. Tremblay, and G. Zar\'and 
for useful discussions and B. Coish for a careful 
reading of the manuscript. K.L.H. 
is supported by NSERC and P.S. by
the Swiss NSF, NCCR, and the EU RTN Spintronics No HPRN-CT-2002-00302.


\begin{thebibliography}{99}
\bibitem{note3}
When the number of electrons in the small dot is even,
we may expect results analogous to \cite{Gramespacher} with a small smearing of the
$log$ singularity.
\end{thebibliography}
\end{document}